# Event-By-Event Fluctuation of Maximum Particle Density in Narrow Pseudo-Rapidity Interval at a Few AGeV/c

Swarnapratim Bhattacharyya

Department of Physics, New Alipore College, L Block, New Alipore, Kolkata 700053, India

Email: swarna_pratim@yahoo.com

## Abstract

A detailed study of event by event fluctuation of maximum particle density of the produced particles in narrow pseudo-rapidity interval in terms of the scaled variance $\omega$ has been carried out for $^{16}$O-AgBr, $^{28}$Si-AgBr and $^{32}$S-AgBr interactions at an incident momentum of 4.5 AGeV/c. For all the interactions the values of scaled variance are found to be greater than zero indicating the presence of strong event by event fluctuation of maximum particle density values in the multiparticle production process.  The event by event fluctuations are found to decrease with the increase of pseudo-rapidity interval. Experimental analysis has been compared with the results obtained from the analysis of events simulated by the Ultra Relativistic Quantum Molecular Dynamics (UrQMD) model. UrQMD model could not replicate the experimental results.

# 1. Introduction

Collision between two heavy ions at relativistic and ultra-relativistic energies produce huge number of various particles and one of the main goals of heavy ion physics is to investigate the particle production mechanism. The produced charged particles have been observed to exhibit number density fluctuations over the considered phase space. Such fluctuation is much larger than the statistical fluctuations arising due to the finiteness of the particle multiplicity produced in a collision. In single particle density distribution presence of large fluctuations in particle density is responsible for the formation of spike like structure. The occurance of such structure is very resourceful to investigate whether the nuclear matter has undergone a quark-hadron phase transition during the collision process [1]. The study of fluctuation and correlation address the fundamental aspects of high energy nucleus-nucleus collisions [2-10]. Investigations involving fluctuations in collisions of heavy nuclei at relativistic energies might serve as a useful tool to explore the particle production mechanism. So the study of nonstatistical fluctuations in relativistic and ultra-relativistic nucleus-nucleus collisions has become a subject of major interest among the particle physicists. To gather any meaningful information about the particle production mechanism, it is therefore important to analyze these fluctuations on event by event basis.

In the recent years studies of event by event fluctuations have gained immense popularity among the scientists. The study of event-by-event fluctuations in high-energy heavy-ion collisions may provide us more information about the multiparticle production dynamics [11-19]. Event-by-event analysis is potentially a powerful technique to the study relativistic heavy-ion collisions, as the magnitude of the fluctuations of various quantities around their mean values is controlled by the dynamics of the system. It is believed that a detailed study of each event produced in high-energy nucleus-nucleus collision, may reveal new phenomena occurred in some rare events. The study of a single event can reveal very different physics than the analysis of averages over a large statistical ensemble. Event-by-event fluctuations may provide us information about the heat capacity [17, 20-22], possible equilibration of the system [23-31] or about the phase transition [22, 32]. The most common way to study event by event fluctuation of any observable is to use the scaled variance [33]. If A is an observable, the scaled variance of A is given by $\omega = \frac{\langle A^2 \rangle - \langle A \rangle^2}{\langle A \rangle}$. If there is no fluctuation, the scaled variance is zero.

One of the most striking signatures of QGP phase transition could be a strong modification in the fluctuations of specific observables measured on an event-by-event basis [18,34]. In principle, any observable that is not globally conserved fluctuates. Although most of these fluctuations are trivial and are of statistical origin, it is important to find out the dynamically relevant event-to-event fluctuation that enables the search for a possible critical point and a first-order coexistence region in the QCD phase diagram [35]. Over the past two decades quite a number of such observables have been suggested for clarifying the evolution of the system formed in heavy-ion collisions. Most commonly measured event-by-event fluctuations in heavy-ion collision experiments are particle ratios, mean transverse momentum and particle multiplicity fluctuations [36-39]. Unfortunately, the volume of the system created in heavy ion collisions cannot be fixed and fluctuates a lot event by event. Within the grand canonical ensemble, the volume cancels if the variance is scaled with the mean multiplicity (an intensive observable), but the dependence on volume fluctuations still remains. As the scaled variance depends on the fluctuations of the volume [40] the problem of reducing this effect becomes very important in fluctuation studies. Effect of conservation laws on event by event fluctuations is also an imporatnt point to be addressed.

The present goal of this study is to investigate event by event fluctuation of maximum particle density of the produced particles in narrow pseudo-rapidity interval in terms of the scaled variance $\omega$ for $^{16}$O-AgBr, $^{28}$Si-AgBr and $^{32}$S-AgBr interactions at an incident momentum of 4.5 AGeV/c. Experimental results have been compared with the results obtained from the analysis of Ultra Relativistic Quantum Molecular Dynamics (UrQMD) model.

## 2. Experimental Details

The present analysis has been carried out with the data obtained from nuclear emulsion track detector. One of the advantages of this detector is that it can record and store the information permanently about the charged particles with different ionizing power over the $4\pi$ geometry. Due to high spatial and good angular resolution (~0.1mrad) emulsion detector is quite acceptable for studying event by event fluctuation of the produced particles when they are distributed in a small phase space interval [41-42].

In order to get the required data for the present analysis NIKFI-BR2 emulsion pellicles of dimension 20cm$\times$10 cm $\times 600\mu$m were irradiated by the $^{16}$O, $^{28}$Si and $^{32}$S beam at 4.5 AGeV/c obtained from the Synchrophasatron at Joint Institute of Nuclear Research (JINR), Dubna, Russia [41-42]. When a projectile collides with the target an interaction or an event occurs. In order to find and interaction or an event we have scanned the emulsion plate along the track of the incident beam starting from the entry point of the beam into the emulsion plate until an interaction occurs. We have also performed the scanning in the backward direction slowly to ensure that the interaction selected by scanning in the forward direction did not include interaction from the secondary tracks of other interactions. Scanning of emulsion plate in both forward and backward direction with two independent observers increases the efficiency of detecting a primary event up to 99%. More details about scanning procedure can be found from our earlier publication [41].

It was Powell [43] who classified the emitted or produced particles in emulsion plate in to four categories, namely the shower particles, the grey particles, the black particles and the projectile fragments. Characteristics of these particles can be found from [41,43]. Nuclear emulsion medium consists of variety of nuclei like H, C, N, O, Ag and Br. It has been pointed out in [41,43] that in emulsion experiment, it is very difficult to measure the charges of the fragments emitted from the target and hence exact identification of the target is not possible. However, we can divide the major constituent elements present in the emulsion into three broad target groups namely hydrogen (H), light nuclei (CNO) and heavy nuclei (AgBr) on the basis of the heavy (black+grey) tracks denoted by $N_h$ as discussed in [41]. For the present analysis we have not considered the events which are found to occur due to collisions of the projectile beam with H and CNO target present in nuclear emulsion. Our analysis has been carried out for the interactions with the AgBr target only. For this study, we have selected 1057 events of $^{16}$O-AgBr, 514 events of $^{28}$Si-AgBr and 434 events of $^{32}$S-AgBr interactions [41-42]. The present analysis has been performed on shower particles only. Average multiplicities of shower tracks of each interaction have been calculated and presented in table 1. We have also calculated the statistical error associated with the average multiplicities of the shower tracks for our experimental events. The errors quoted in table 1 are a subjective estimate based on the sample standard deviation.

The emission angle ($\theta$) was measured with respect to the direction of the incident beam for each track by taking readings of the coordinates of the interaction point ($X_0$, $Y_0$, $Z_0$), coordinates ($X_1$, $Y_1$, $Z_1$) at the end of the each secondary track and coordinates ($X_i$, $Y_i$, $Z_i$) of a point on the incident beam. In case of shower particle multiplicity distribution the phase space variable used is pseudo-rapidity $\eta$. The relation $\eta = -\ln \tan \frac{\theta}{2}$ relates the variable $\eta$ with the emission angle $\theta$. In an emulsion experiment, the pseudo-rapidity is a convenient choice for the basic variable in terms of which the experimental data can be analyzed.

In order to extract event-by-event dynamical fluctuations, one has to remove trivial effects, which also lead to fluctuations of the particle multiplicity measured on event-by-event basis. The major effect in this respect is the variation of the impact parameter. In emulsion detector it is impossible to know the impact parameter. However on the basis of the the total charge or sum of the charges of the non-interacting projectile fragments, it may be possible to select the central events. In a recent paper [44] we have studied the multiplicity distribution of shower particles in central collisions using emulsion detector. From that paper we may said that we have 20.94% central events in $^{16}$O-AgBr interactions, 13.70% central events for $^{28}$Si-AgBr and 7.22% central events in $^{32}$S-AgBr interactions [44]. On the other hand, selection of peripheral events depends on the value of $N_h$. Events having $N_h$=0 are designated as peripheral events [45]. As we are dealing with AgBr events ($N_h$>8) there is no peripheral events in our analysis. So we are dealing with central and quasi central collisions. However the exact range of centrality can not be determined in emulsion detector.

### 3. Analysis and Results

Before going into the details of our analysis it will be convenient for the readers to have a look into the pseudo-rapidity distribution of the data sample. Figure 1(a)-1(c) represent the pseudo-rapidity distribution of shower particles for $^{16}$O-AgBr, $^{28}$Si-AgBr and $^{32}$S-AgBr interactions at an incident momentum of 4.5 AGeV/c.

In order to calculate the event by event fluctuation of maximum particle density in narrow pseudo-rapidity interval for $^{16}$O-AgBr, $^{28}$Si-AgBr and $^{32}$S-AgBr interactions at an incident momentum of 4.5 AGeV/c, we have first calculated the maximum density of charged particles following the method described by E.K. Sarkisyan [46] and D.Ghosh [47]. According to them maximum density of particles is defined as $\rho_{max} = \frac{\Delta N_{max}}{\Delta \eta}$..............(1).

Where $\Delta N_{max}$ is the maximum number of particles within the interval $\Delta \eta$ of each event. We have calculated the values of $\rho_{max}$ of each event for five different $\Delta \eta$ intervals selected as $\Delta \eta = 0.1, 0.3, 0.5, 0.8$ and $1.0$ for all the three interactions. The quantification of event by event fluctuations of the maximum particle density has been performed with the variable $\omega$, called the scaled variance, as discussed earlier, so that here $\omega = \frac{\langle \rho_{max}^2 \rangle - \langle \rho_{max} \rangle^2}{\langle \rho_{max} \rangle}$. The averaging has been done over all the events for a particular interaction. We have calculated the value of the scaled variance $\omega$ for maximum particle density fluctuation and presented the calculated values in table 2 for $^{16}$O-AgBr, $^{28}$Si-AgBr and $^{32}$S-AgBr interactions. Errors associated with the event by event fluctuations of the maximum particle density values are the statistical errors.

From the table it may be noted that the values of the variable $\omega$ decreases as the pseudo-rapidity window size increases. Table 2 signifies that for all the interactions the values of scaled variance are found to be greater than zero indicating the presence of strong event by event fluctuation of maximum particle density in the multiparticle production process. Moreover for heavier projectile the value of the variable $\omega$ is higher signifying stronger event by event fluctuation for heavier projectile. The occurrence of event by event fluctuations may be attributed to the fact that particles were produced in a correlated manner. The variation of the variable $\omega$ with the pseudo-rapidity interval $\Delta\eta$ has been presented in figure 2-figure 4 for $^{16}$O-AgBr, $^{28}$Si-AgBr and $^{32}$S-AgBr interactions respectively.

The experimental results have been compared with those obtained by analyzing events generated by the Ultra Relativistic Quantum Molecular Dynamics (UrQMD) model. UrQMD model is a hadronic transport model based on the covariant propagation of all the hadrons on the classical trajectories in combination with stochastic binary scattering, color string formation and resonance decay. This model can be used in the entire available range of energies from the Bevalac region to RHIC and LHC to simulate the nucleus-nucleus collisions. For more details about this model, readers are requested to consult [48-49]. We have generated a large sample of events using the UrQMD code (UrQMD 3.3p1) for $^{16}$O-AgBr, $^{28}$Si-AgBr and $^{32}$S-AgBr interactions in the pseudo-rapidity $(\eta)$ space [50]. Taking Ag and Br nuclei as target, events were generated separately for each target. According to the proportional abundance of Ag and Br nuclei present in the nuclear emulsion, the generated events were mixed with each other in order to get the desired UrQMD data sample [42]. All the charged mesons produced in the UrQMD simulation were considered for the analysis of event by event fluctuation of maximum particle density. We have also calculated the average multiplicities of the shower tracks for all the three interactions in case of the UrQMD data sample. Average multiplicities of the shower tracks in case of UrQMD data sample have been presented in table 1 along with the average multiplicity values of shower particles in the case of the experimental events [50]. Table 1 shows that the average multiplicities of the shower tracks for the UrQMD events are comparable with those of the experimental values for all the interactions. The pseudo-rapidity distribution of the UrQMD-model–generated data sample have been presented in figs. 1(a)–1(c) along with the experimental pseudo-rapidity distribution.

We have calculated the event by event fluctuation variable for maximum particle density in $^{16}$O-AgBr, $^{28}$Si-AgBr and $^{32}$S-AgBr interactions for $\Delta\eta = 0.1, 0.3, 0.5, 0.8$ and $1.0$ for the UrQMD simulated events. Calculated values of the variable $\omega$ for five different $\Delta\eta$ intervals have been presented in table 2. The values of the variable $\omega$ quantifying the event by event fluctuation of maximum particle density are greater than zero for all the interactions in case of UrQMD simulated data as evident from table 2. It may be noted from the table that like the experimental analysis, in case of UrQMD study also the event by event fluctuation variable decrease with the increase of $\Delta\eta$. From the table it can be concluded that the values of variable $\omega$ are significantly lower than the corresponding experimental values. Thus UrQMD model can not reproduce the experimental results. The variation of variable $\omega$ with the pseudo-rapidity interval $\Delta\eta$ has been depicted in figure 2-figure 4 for $^{16}$O-AgBr, $^{28}$Si-AgBr and $^{32}$S-AgBr interactions respectively for the UrQMD simulated data.

Before we conclude, let us discuss the effect of systematic errors on our analysis. In nuclear emulsion detector sources of systematic errors are the scanning procedure, fading of tracks and the presence of background contaminations. It has been discussed in our earlier papers that total contribution of systematic errors coming out from the scanning procedure, fading of tracks, and presence of background contaminations is around (1-2)% [42,45,52]. It has been discussed in [50] that the shower particles are mostly pions (more than 90%) with a small proportion (less than 10%) of kaons and hyperons among them. The presence of K-mesons, hyperons and any other mesons among the pions are treated as contaminations. As nuclear emulsion track detector cannot distinguish between pions and other mesons or hyperons, one possible source of systematic uncertainty is the presence of such contaminations among the shower particles. We have calculated the maximum systematic uncertainty while dealing with shower particles for $^{16}$O-AgBr, $^{28}$Si-AgBr and $^{32}$S-AgBr interactions in [51]. The contribution to the systematic errors due to the presence of other mesons and hyperons with the pions in the shower particles has been calculated [51]. Total contribution of systematic errors in our analysis for $^{16}$O-AgBr, $^{28}$Si-AgBr and $^{32}$S-AgBr interactions at an incident momentum of 4.5 AGeV/c are 9.60%, 10.22% and 10.56% respectively [42].

## 4. Conclusions

We have presented a detailed study of event by event fluctuation of maximum particle density of the produced particles in narrow pseudo-rapidity interval $\Delta\eta = 0.1, 0.3, 0.5, 0.8$ and $1.0$ in terms of the scaled variance $\omega$ for $^{16}$O-AgBr, $^{28}$Si-AgBr and $^{32}$S-AgBr interactions at an incident momentum of 4.5 AGeV/c. For all the interactions the values of scaled variance are found to be greater than zero indicating the presence of strong event by event fluctuation of maximum particle density values in the multiparticle production process. The event by event fluctuations are found to decrease with the increase of pseudo-rapidity interval. These fluctuations are found to exhibit strong projectile dependence. Presence of dynamical correlation during the particle production process is the physical origin of event by event fluctuations of maximum particle density. Experimental analysis has been compared with the results obtained from the analysis of Ultra Relativistic Quantum Molecular Dynamics (UrQMD) model. UrQMD model could not replicate the experimental results. This is the first ever report of event by event fluctuation of maximum particle density of the produced particles in high energy nucleus-nucleus collisions at a few AGeV/c. The study is interesting and deserves attention.

**Table 1**

| Interactions at 4.5AGeV/c | Average Multiplicity of shower particles | |
|---|---|---|
| | Experimental | UrQMD |
| $^{16}$O-AgBr | 18.05±0.22 | 17.79±0.21 |
| $^{28}$Si- AgBr | 23.62±0.21 | 27.55±0.22 |
| $^{32}$S- AgBr | 28.04±0.14 | 30.84±0.17 |

Table 1 represents the average shower particle multiplicities for $^{16}$O-AgBr, $^{28}$Si-AgBr and $^{32}$S-AgBr interactions at 4.5AGeV/c in case of experimental and UrQMD data. Statistical error associated with average multiplicity has been mentioned.

**Table 2**

| Interactions at 4.5AGeV/c | Calculated values of $\omega$, the measure of event by event fluctuation of maximum particle density in different narrow pseudo-rapidity interval | | |
|---|---|---|---|
| | pseudo-rapidity interval $\Delta\eta$ | Experimental Value | UrQMD simulated Value |
| $^{16}$O-AgBr | 0.1 | 4.91±.05 | 2.67±.02 |
| | 0.3 | 3.85±.06 | 1.51±.02 |
| | 0.5 | 3.42±.07 | 1.20±.02 |
| | 0.8 | 3.12±.08 | 1.03±.02 |
| | 1.0 | 2.99±.09 | 0.96±.03 |
| $^{28}$Si-AgBr | 0.1 | 7.29±.06 | 2.76±.02 |
| | 0.3 | 5.50±.07 | 1.57±.03 |
| | 0.5 | 5.21±.08 | 1.33±.03 |
| | 0.8 | 4.69±.09 | 1.15±.03 |
| | 1.0 | 4.51±.09 | 1.07±.04 |
| $^{32}$S-AgBr | 0.1 | 8.06±.07 | 2.76±.02 |
| | 0.3 | 5.96±.09 | 1.57±.05 |
| | 0.5 | 5.70±.11 | 1.33±.04 |
| | 0.8 | 5.19±.12 | 1.15±.05 |
| | 1.0 | 4.94±.13 | 1.08±.05 |

Table 2 represents the values of event by event fluctuation of maximum particle density in narrow pseudo-rapidity interval for $^{16}$O-AgBr, $^{28}$Si-AgBr and

$^{32}$S-AgBr interactions at an incident momentum of 4.5 AGeV/c in case of experimental and UrQMD simulated data. Errors shown in the table are statistical errors only.

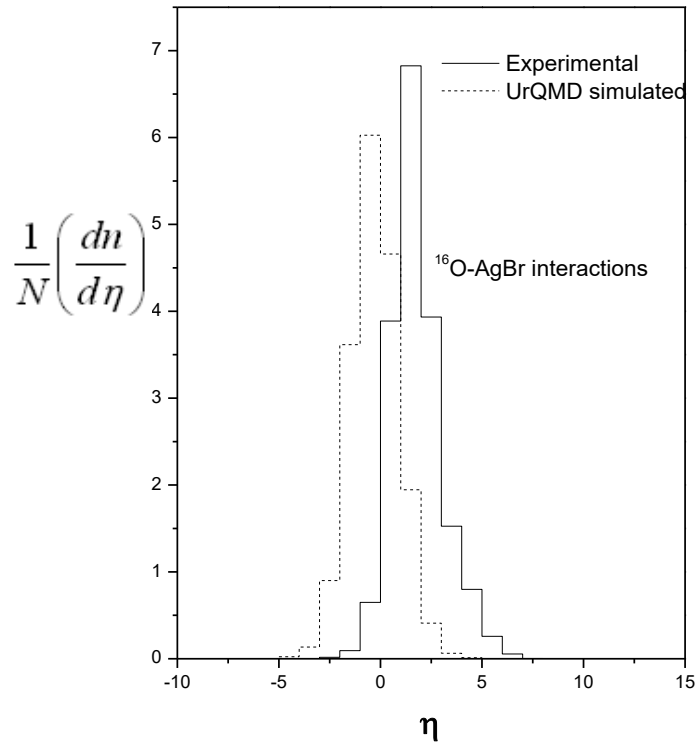

Fig 1(a) Pseudo-rapidity Distribution of shower particles for the experimental and UrQMD events for $^{16}$O-AgBr interactions

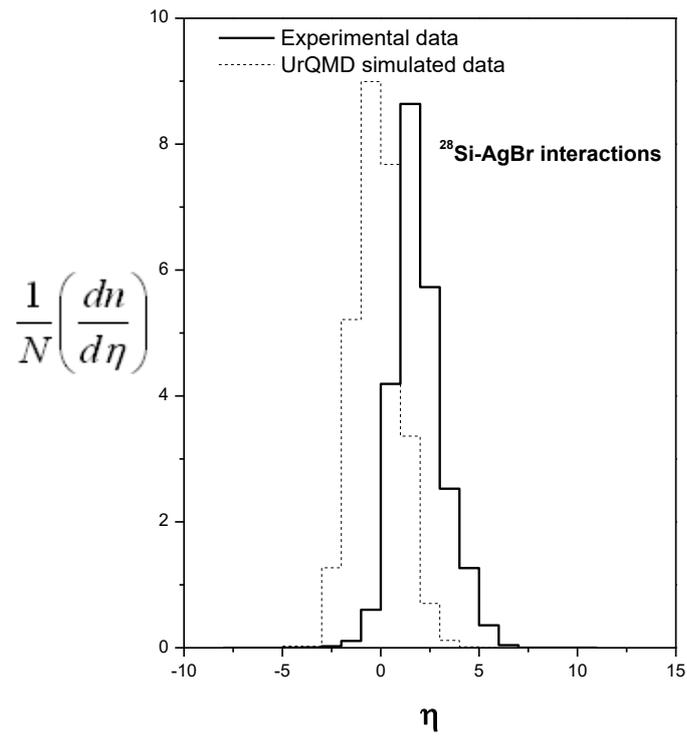

Fig 1(b) Pseudo-rapidity Distribution of shower particles for the experimental and UrQMD events for $^{28}$Si-AgBr interactions

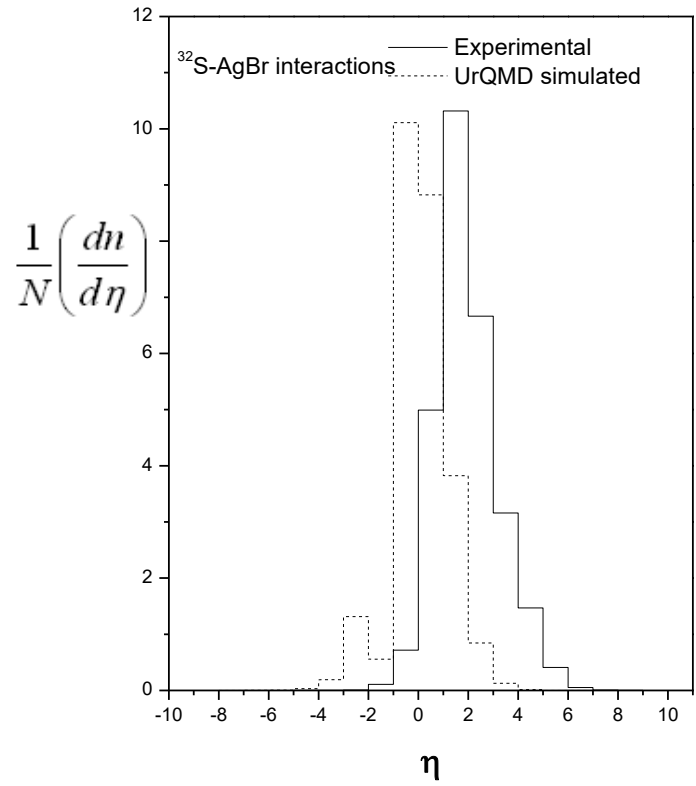

Fig 1(c) Pseudo-rapidity Distribution of shower particles for the experimental and UrQMD events for $^{32}$S-AgBr interactions

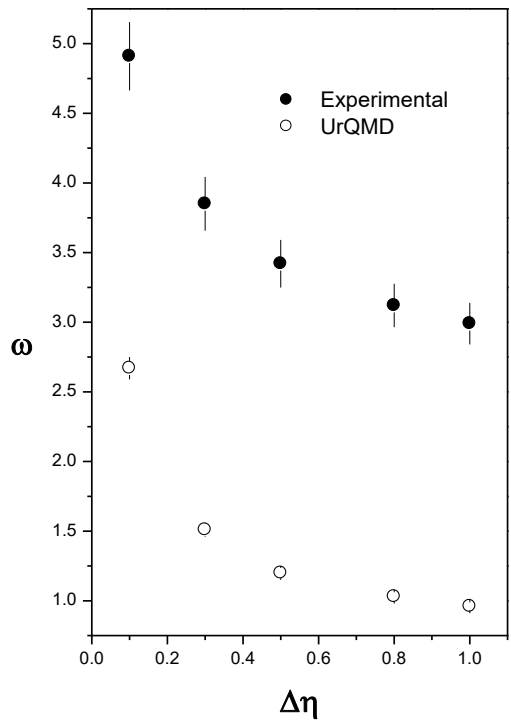

Figure 2: Variation of event by event fluctuation of maximum particle density with respect to the narrow pseudo-rapidity interval in $^{16}$O-AgBr interactions at 4.5 AGeV/c.

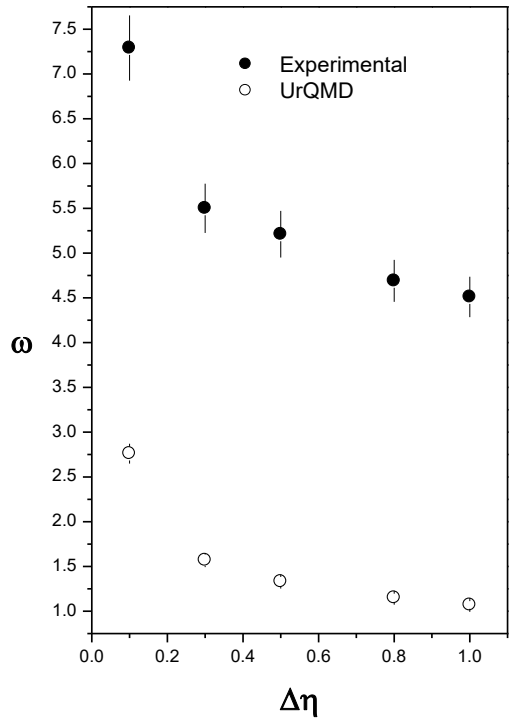

Figure 3: Variation of event by event fluctuation of maximum particle density with respect to the narrow pseudo-rapidity interval in $^{28}$Si-AgBr interactions at 4.5 AGeV/c.

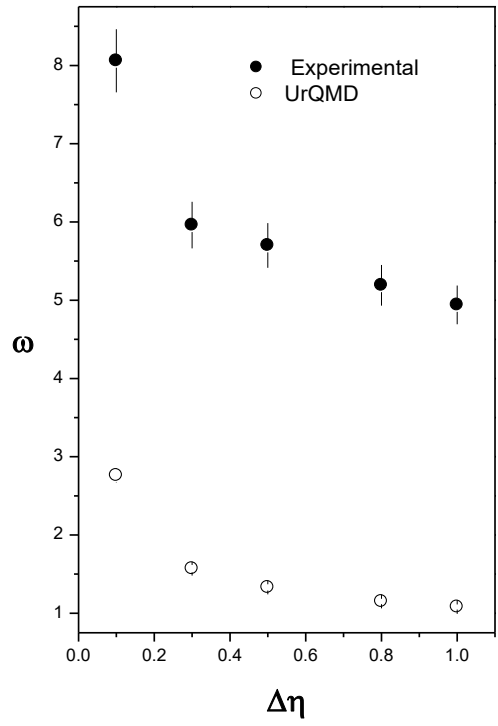

Figure 4: Variation of event by event fluctuation of maximum particle density with respect to the pseudo-rapidity interval in $^{32}$S-AgBr interactions at 4.5 AGeV/c.

## Acknowledgement

The author is grateful to Prof. Pavel Zarubin, JINR, Dubna, Russia for providing the required emulsion data. Dr. Bhattacharyya also acknowledges Prof. Dipak Ghosh, Department of Physics, Jadavpur University and Prof. Argha Deb Department of Physics, Jadavpur University, for their inspiration in the preparation of this manuscript.